\documentclass[reprint,amsmath,amssymb]{revtex4-2}
\usepackage{amsmath}
\usepackage{array}
\usepackage{amssymb}
\usepackage{subfig}
\setcitestyle{numbers,square}
\usepackage[colorlinks=true, allcolors=blue]{hyperref}
\usepackage{threeparttable}
\usepackage{graphicx}
\usepackage{dcolumn}
\usepackage{bm}
\usepackage{cleveref}
\usepackage{caption}
\usepackage{xcolor}
\usepackage{ragged2e}
\usepackage{multirow}

\usepackage{xcolor}
\usepackage[normalem]{ulem}

\begin{document}

\title{Physical Design of Positron Transport System for Muonium-to-Antimuonium Conversion Experiment}

\author{Guihao Lu}

\author{Shihan Zhao}

\author{Siyuan Chen}
\author{Jian Tang}
\email{tangjian5@mail.sysu.edu.cn}
\affiliation{School of Physics, Sun Yat-sen University, Guangzhou, 510275, China}
\affiliation{Platform for Muon Science and Technology, Sun Yat-sen University, Guangzhou 510275, China}

\begin{abstract}

Muonium-to-Antimuonium Conversion Experiment (MACE) aims to find the charged lepton flavor violation (cLFV) process. A key component of MACE is the positron transport system (PTS) to collect and transport atomic positrons from antimuonium decays, which consists of an electrostatic accelerator and a solenoid beamline.  Through field simulations in \textsc{COMSOL} and particle transport simulations based on \textsc{Geant4}, the PTS can achieve a geometric acceptance of the signal at 65.81(4)\% along with a position resolution of 88(1)~$\mu$m~$\times$~102(1)~$\mu$m. The system achieves 322.4(1)~ns transit time with a spread of 6.9(1)~ns, which allows for a TOF-based rejection of internal conversion backgrounds by a factor of $10^{-7}$. These promising results pave the way for new-physics signal identifications and background rejections in MACE and offer a novel paradigm for internal transport system in high-intensity frontiers.

\end{abstract}

\maketitle

\section{Introduction}
\label{introduction}
The Standard Model (SM) remains incomplete. Neutrino oscillations demonstrate that the lepton-flavor-conserving symmetry $U(1)_{L_e} \times U(1)_{L_\mu} \times U(1)_{L_\tau}$ is broken in the neutral lepton sector. It is still a mystery whether there is an observable process of charged lepton flavor violation (cLFV)~\cite{Petcov:1976ff,Bilenky:1977du,Lee:1977tib}. This motivates searches for new physics beyond the Standard Model (BSM) in the charged lepton sector~\cite{Calibbi:2017uvl,Petrov:2022wau,Ardu:2022sbt}. Muons provide uniquely sensitive BSM probes~\cite{Kuno:1999jp,Lindner:2016bgg}, driving experiments including COMET~\cite{COMET:2018auw}, Mu2e~\cite{Mu2e:2014fns}, MEG-II~\cite{MEGII:2023fog,MEGII:2025gzr}, Mu3e~\cite{Mu3e:2020gyw}, and MACE~\cite{Bai:2024skk}. MACE aims at searching for muonium-to-antimuonium conversion, a lepton flavor violation process by two units, accessing rich and distinct BSM physics~\cite{Fukuyama:2021iyw,Calibbi:2024rcm, Heeck:2024uiz,Heeck:2025jfs}.

MACE utilizes high-intensity surface muons from the CiADS (China Initiative Accelerator Driven Subcritical System)~\cite{He:2023izb}. The CiADS facility, currently under construction, features a superconducting linac designed to deliver a proton beam with an energy of 500~MeV (upgradable to 600~MeV) and a high current of 5~mA in continuous-wave (CW) mode~\cite{ Wang:2019ddc}. Such a high-power proton driver (up to 2.5–3.0~MW) provides a unique opportunity for producing world-class high-intensity muon beams. MACE prefers surface muons due to their low momentum and low energy spread. When a $\pi^+$ decays at rest ($\pi^+\to \mu^+ + \nu_{\mu}$), the resulting $\mu$ is emitted with a monochromatic momentum of 29.8 MeV/c~\cite{Cai:2023caf}, within a thin layer (the surface) of the production target. Specifically, at the CiADS muon source, a novel free-surface liquid lithium jet target is proposed to handle the unprecedented heat load from the 5~mA proton beam.

Signal events of MACE are identified by measuring the antimuonium decay products. In this experiment, the surface muons stopped in the target interact with atomic electrons to form muonium ($\text{M}=\mu^+e^-$), a hydrogen-like bound state~\cite{Zhao:2023plv}. The observation of muonium-to-antimuonium conversion ($\mu^+e^- \rightarrow \mu^-e^+$) would demonstrate a lepton flavor violation by two units. The $\mu^-$ in antimuonium undergoes a weak decay into a Michel electron (energy up to 52.8~MeV) and two neutrinos~\cite{Czarnecki:1999yj}. At the same time, the atomic positron is released from the bound state to a free positron with a mean energy of 13.5~eV. The signature, a high-energy electron and a low-energy positron, motivates the detector and geometry optimization.

A background event involves a high-energy electron and a low-energy positron. In the Standard Model, five-body decays of a muon $\mu^+ \to e^+ e^- e^+ \nu_e \bar{\nu}_{\mu}$ contribute to such a background event. This process has a branching ratio of $3.4 \times 10^{-5}$ (with a transverse momentum cut of $p_\text{T} > 17~\text{MeV}/c$)~\cite{SINDRUM:1985vbg}. The internal conversion (IC) decay can mimic signal events if these three conditions are satisfied: (1) the electron is detected; (2) one of the positrons has very low momentum; and (3) the second positron remains undetected. With a high-intensity muon beam of $10^8$--$10^{10}~\mu^+/s$, this process can contribute significantly to the background. Since the positron from the internal conversion decay of a muon (IC muon decay) has a high energy, applying an energy selection design on the positron can significantly suppress this background.

The current best upper limit on the muonium-to-antimuonium conversion probability, $P_{\text{M}\overline{\text{M}}}\leq 8.2\times10 ^{-11}$~(90\% C.L.), was established by the MACS experiment at PSI in 1998~\cite{Willmann_1999}. To improve the sensitivity by more than two orders of magnitude, MACE will implement a novel Positron Transport System designed for more rigorous background rejection. Inspired by the transport system of COMET~\cite{Arimoto:2024zbd} and Mu2e~\cite{Ambrosio:2013/07/16fca}, we adopt a symmetric S-shaped solenoid configuration.

The role of the PTS in MACE differs from that of the transport solenoid in Mu2e. The Mu2e transport solenoid is designed to deliver muons to the detector region~\cite{Lynch:2017tsx}. In contrast, the PTS in MACE transports positrons from the target region to the positron detection system, which serves as a component of the whole detector system. Despite the momentum selection similar to Mu2e, the PTS allows for signal reconstruction and efficient background rejection. For these purposes, an electrostatic acceleration system is coupled within the beam pipe at the center of the detector, forming a distinctive component. Together with the magnetic field characteristics of the S-shaped solenoid, a collimator is introduced to further suppress the backgrounds. The PTS in MACE addresses the unique requirements for low-energy signal transmission and selection, which offers a novel paradigm for internal transport systems in high-intensity frontiers.

This paper is organized as follows. Section~\ref{sec:PTS} introduces the layout, parameters, and functions of the components in the solenoid beamline, including numerical simulation methods and optimization strategies. Section~\ref{sec:results} presents the transmission simulated by \textsc{Geant4} with the field calculations imported from COMSOL and demonstrates the excellent total selection efficiency of the signals and significant suppression of IC muon decay backgrounds. Finally, we draw our conclusions in Section~\ref{sec:conclusion}.

\section{Positron Transport System Design}
\label{sec:PTS}

\begin{figure*}[ht]
    \centering
    \includegraphics[width=0.9\linewidth]{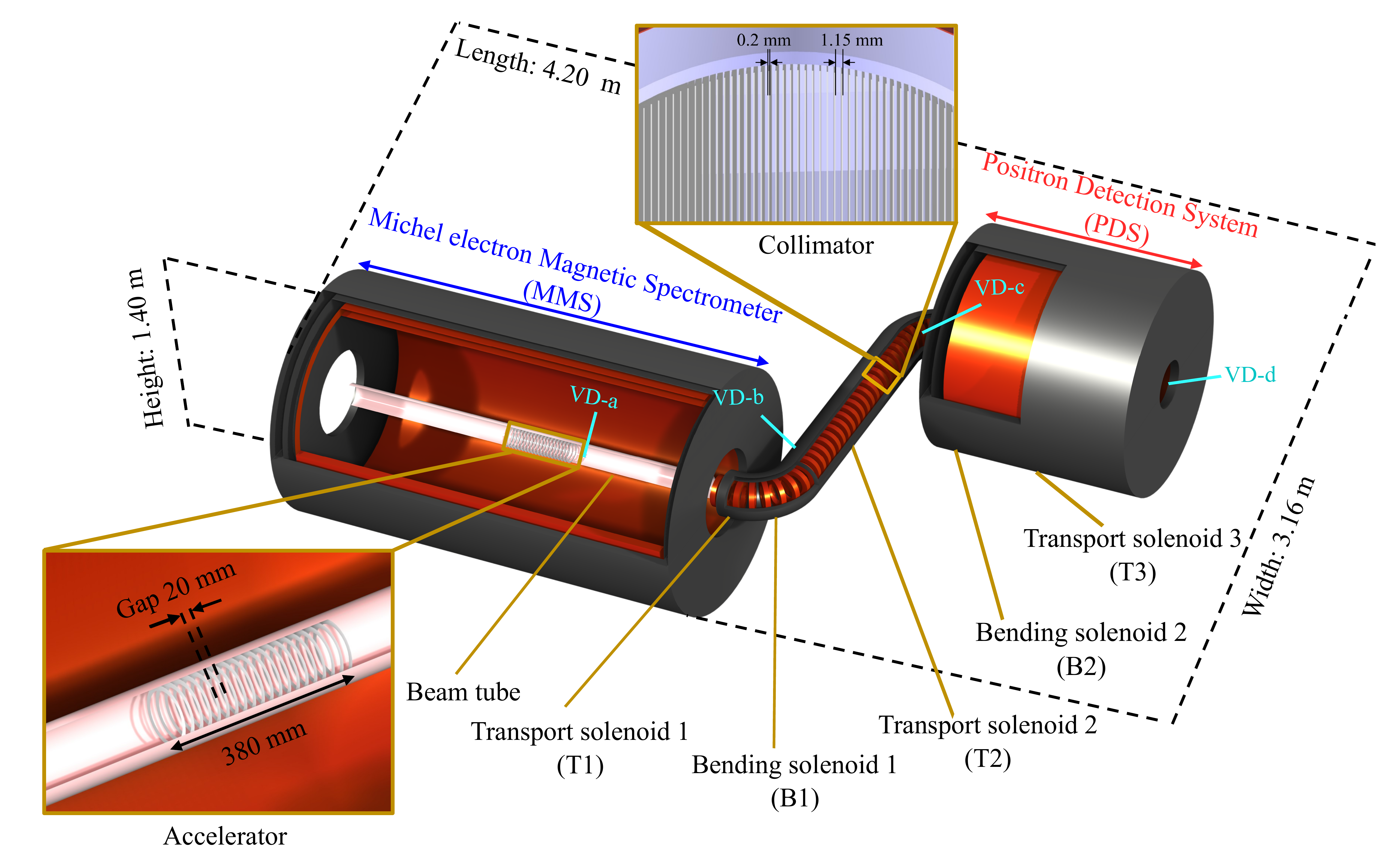}
    \caption{A schematic figure of the positron transport system. The outer grey shell represents the iron yoke, while the copper-colored interior denotes the solenoid. The key components and their corresponding locations have been labeled in the diagram. B2 and T3 point to their positions in perspective. VDs are virtual detectors with different positions in the simulation. }
    \label{fig:PTS_Overview}
\end{figure*}

The general architecture of the PTS is illustrated in Figure~\ref{fig:PTS_Overview}. Its function begins with surface muons injected into a target at the center of the accelerator, leading to the formation of muonium. In the event of conversion to antimuonium and its subsequent decay, a low-energy atomic positron is produced. This positron is immediately accelerated and then guided by a solenoid beamline. The
beamline consists of the Michel electron Magnetic Spectrometer (MMS) solenoid, the S-shaped transport solenoid housing a collimator, and the Positron Detection System (PDS) solenoid. The entire system is engineered to transport these signal positrons to hit the Microchannel plate (MCP)~\cite{Chen:2024jmg} located in the middle of PDS with high geometric acceptance, enabling both precise vertex reconstruction and strong background suppression.

\subsection{Solenoid system}

Achieving high spatial resolution and geometric acceptance in positron transport requires a consistent longitudinal magnetic field throughout the beamline. Thus, the signal of low-energy positrons can be transported closely along the magnetic field lines parallel to the central axis, thereby facilitating subsequent signal reconstruction. However, the fringe magnetic fields produced by the MMS and PDS solenoids are comparable to the magnetic fields in the B1, T2, and B2 regions, resulting in distortion of the transport solenoid field. Due to the smaller aperture of the transport solenoid compared to that of the MMS and PDS solenoid, this interference is naturally mitigated by the use of iron yokes. Although magnetic field overlap at the interfaces is unavoidable, fringe effects can be effectively minimized by fine-tuning the currents in the T1, B1, T2, and B2 coils.

As discussed in references~\cite{Hou:1995np,Fukuyama:2023drl}, muonium conversion induced by effective couplings $(V\pm A) \times (V\pm A)$ and $(S\pm P) \times (S\pm P)$ is significantly suppressed when the magnetic field reaches 1~T or higher. Therefore, following an assessment of the physics performance in MACE, the magnetic field strength of the solenoid system is set to 0.1~T.

To achieve the required 0.1~T longitudinal magnetic field and ensure high spatial resolution, the PTS employs a specialized solenoid beamline consisting of three key components: (i) a cylindrical MMS solenoid (2400~mm length × 700~mm radius) surrounded by 50-mm-thick iron yoke, providing both magnetic field for MMS detector and initial positron transport; (ii) an S-shaped transport solenoid composed of three straight segments (T1/T3: 150~mm; T2: 1314.5~mm with integrated collimator) and two counter-rotating 90° toroids (250~mm bend radius), all with identical coil geometry (30~mm length, 120/180~mm inner/outer diameters) and encased in 30-mm-thick iron yoke; and (iii) a cylindrical PDS solenoid (1200~mm × 650~mm) with a 50-mm-thick iron yoke for containing the final-stage positrons. 

The S-shaped design of the transport solenoid helps to suppress the background. Signal positrons, which have low momentum, tend to follow the magnetic field lines during transport. In contrast, backgrounds with high momentum are more likely to hit the beam pipe at the S-shaped bends, effectively serving as a momentum filter. When positrons reach the MCP, their transverse positions and arrival time are recorded. By applying a position-mapping algorithm, the decay position of the parent particle in the target region can be reconstructed. In this way, the PTS links the spatial and timing information of events between the PDS and MMS, enabling the identification of antimuonium events.

Specifically, the adoption of the symmetric S-shaped configuration, as opposed to a single-bend topology, is essential for correcting beam trajectories. When traversing a single curved solenoid section at 90$^\circ$, charged particles experience a transverse drift~\cite{Arimoto:2024zbd}:
\begin{equation}
D = \frac{\pi}{2}\cdot\frac{p}{q B}.
\end{equation}
where $B$ is the magnetic field on the axis, and $p$, $q$ are the particle momentum and electric charge, respectively. On one hand, the employment of a symmetric S-shaped arrangement allows for the effective cancellation of drift, which leads to higher precision in signal reconstruction. On the other hand, bending magnetic fields are prone to fringe-field effects, which lead to particle divergence and consequently degrade the reconstruction precision. The symmetric S-shaped configuration mitigates these fringe-field effects. Eventually, employing a symmetric S-shaped design therefore improves the reconstruction precision.

\begin{figure}[h!]
\centering 
\subfloat[The simulated distribution of an overall magnetic field strength in the cross section of PTS with magnetic field lines. (the unit of color-bar is T).]{\raisebox{30pt}{\includegraphics[width=0.9\linewidth]{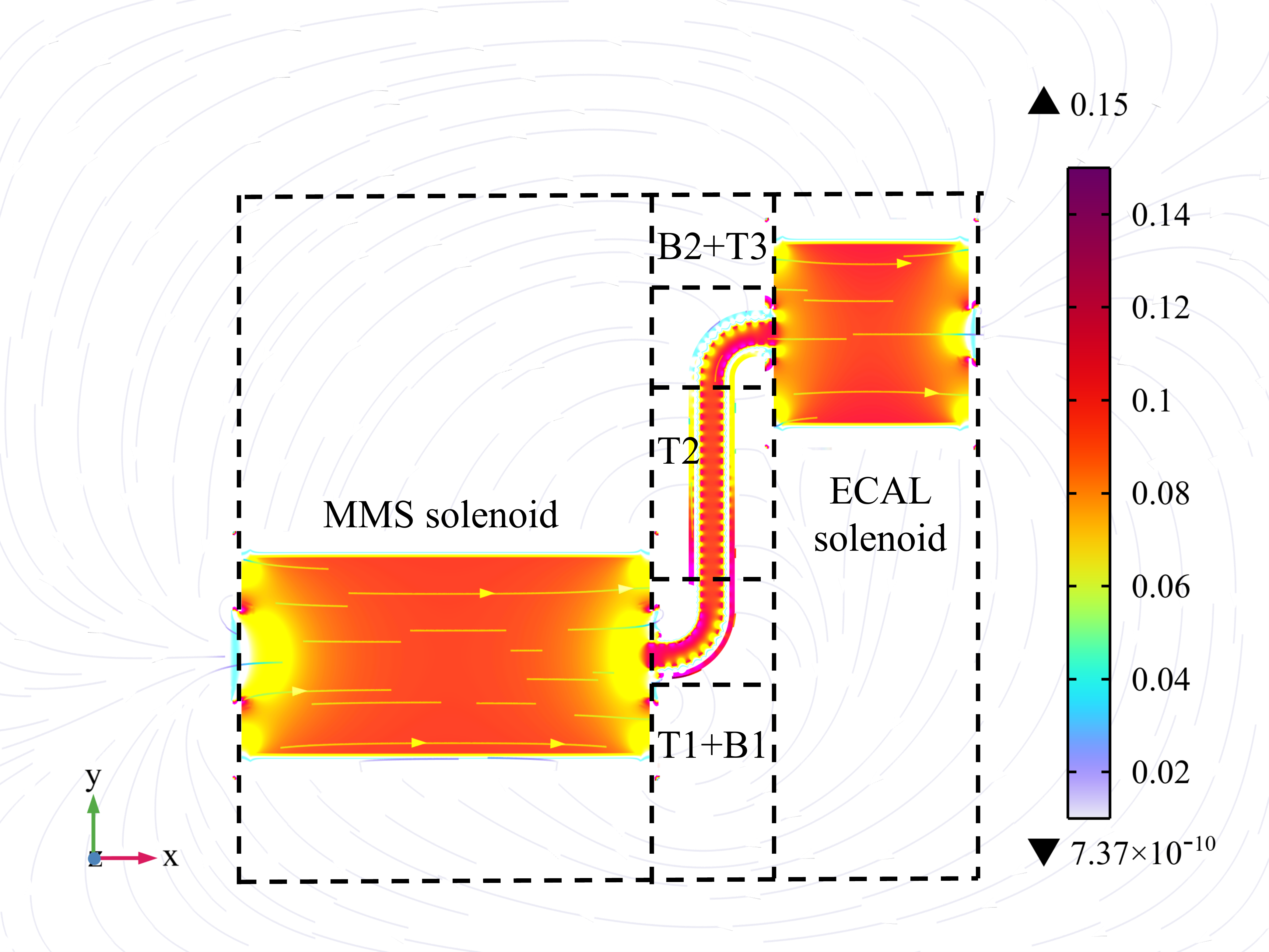}}}
\\
\subfloat[Distribution of the magnetic field along the axis of PTS with the corresponding solenoid regions, extreme values, and the ideal target magnetic field.]{\includegraphics[width=0.9\linewidth]{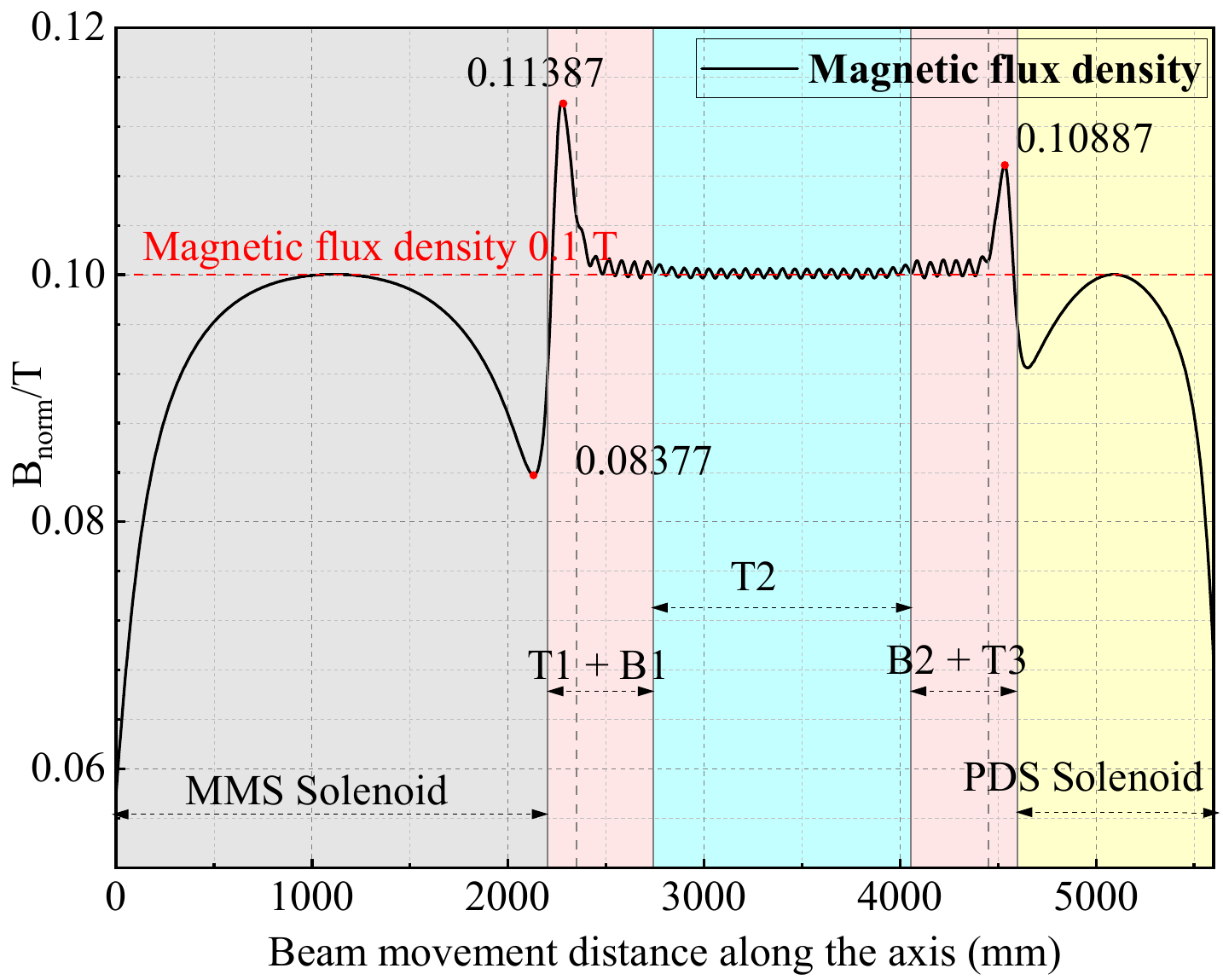}}
\caption{Overview of the magnetic field of the PTS system.}
\label{fig:PTS_MEG}
\end{figure}

The electromagnetic field is calculated using the electromagnetic module of \textsc{COMSOL}~\cite{multiphysics1998introduction}. The magnetic field is reasonably uniform as illustrated in Fig.~\ref{fig:PTS_MEG}. The field can be tuned to nearly 0.1~T by adjusting the currents in the individual solenoid components.
As listed in Table~\ref{tab:Current}, the currents in different solenoids are separately tunable to achieve an almost uniform field.

The transport solenoid is composed of segmented coils, allowing current fine-tuning for each section to compensate for fringe effects and field overlaps. In the figure, the slight increase in field strength near the end of the MMS solenoid, reaching up to 0.114~T, is caused by the overlap between the T1 section and the MMS solenoid yoke. A similar effect is seen at the T3 end, but the PDS solenoid is smaller in size and exhibits a less significant edge effect. In general, the field deviation due to edge effects remains well under control with the uncertainty smaller than 12\% of the required magnetic field after optimization, so that the field performance is sufficient to ensure reliable signal transmission and accurate vertex reconstruction.

\begin{table}[ht]
\centering
\caption{The current in each component after optimization.}
\label{tab:Current}
\begin{threeparttable}
\begin{tabular}{cccc}
\hline\hline
\multicolumn{2}{c}{Component} &  Current (A) & Number of coils \\ \hline
MMS Solenoid                 &                      & 177107.0    & 1        \\ \hline
\multirow{3}{*}{TS Solenoid} & T1/T3                & 4154.5      & 2~$\times$~2      \\ \cline{2-4} 
                             & B1/B2                & 5235.0      & 6~$\times$~2      \\ \cline{2-4} 
                             & T2                   & 4895.4      & 21       \\ \hline
PDS solenoid                & \multicolumn{1}{l}{} & 85512.6     & 1        \\
\hline\hline
\end{tabular}

\end{threeparttable}
\end{table}

\subsection{Collimator}
\label{sec:collimator}

One of the dominant backgrounds in MACE arises from the IC muon decay process, which produces positrons and electrons with significantly higher kinetic energy than the signal positrons. While the signal positrons have a mean energy of 13.5~eV, background positrons typically occupy the MeV energy region. Though the S-shaped transport solenoid effectively rejects high-momentum particles, a minor fraction of the background in the lower energy tails (keV region) contributes significantly to the overall background rate due to the high intensity of $\mu^+$. This challenge needs additional momentum filtering. When these higher-momentum background particles enter the first bending solenoid, the magnetic field induces an exchange between their longitudinal and transverse momentum components, resulting in a large Larmor radius as they travel in the PTS. With the characteristic of the bending magnetic field, the PTS employs a dedicated collimator to suppress these backgrounds by imposing a strict cut on the transverse momentum of the transmitted particles. The resulting momentum selection is quantitatively illustrated in Section~\ref{sec:results-B}.

The PTS collimator is located at the center of the T2 segment of the magnet. As illustrated in Fig.~\ref{fig:PTS_Overview}, the collimator consists of bronze sheets arranged parallel to the beam direction, blocking particles with high transverse momentum while allowing most of the signal positrons to pass. 

Precise momentum selection is achieved by adjusting the pitch ($l$) between the bronze sheets, which determines the transverse momentum ($p_{\mathrm{T}}$) cutoff required to distinguish the signal from the background. By optimizing the signal-to-noise ratio, we find an optimal pitch of approximately 1.15 mm between the bronze sheets, corresponding to a transverse momentum cut of 14~keV/$c$. However, the thickness of bronze sheets ($d$) would reduce the signal acceptance. Considering the mechanical strength of the collimator, the thickness of a single bronze sheet ($d$) is set to $0.2~\text{mm}$.

The Larmor radius ($r$) of the signal positrons in the bending magnetic field is estimated to be $0.1~\text{mm}$, later supported by the signal spatial resolution in Section~\ref{sec:results-a}. Utilizing the plate pitch $l=1.15 \text{ mm}$, the signal acceptance ($\varepsilon$) of the collimator can be estimated by formula:
\begin{equation}
\varepsilon = 1 - \frac{d + 2r}{l}.
\label{eq:collimator}
\end{equation}
Plugging in values of the geometric parameters, we immediately obtain the estimated acceptance at 65\%. 

The collimator is designed with a total length of $0.5~\text{m}$. Given the magnetic field of 0.1~T, the cyclotron frequency for non-relativistic electrons is approximately $2.8~\text{GHz}$. The particles in such a high cyclotron frequency are guaranteed to complete multiple turns within the collimator, fully exploiting its momentum-selection capability. It is noted that the performance of such a high-precision device is extremely sensitive to the alignment. For the current design with a 1.15~mm pitch and 500~mm length, a slight angular misalignment of merely 0.13$^\circ$ relative to the beam axis would result in a complete loss of signals. The length will be further optimized under the constraints from future engineering design. Consequently, a detailed study of misalignment and its impact on the background level and the signal efficiency will be a focus of future engineering design and should be validated by the prototype demonstration.

\begin{figure}[h!]
\centering 
\subfloat[A schematic diagram illustrating the electric field distribution of the electrostatic accelerator (the unit of color-bar is E/m).]{\includegraphics[width=0.9\linewidth]{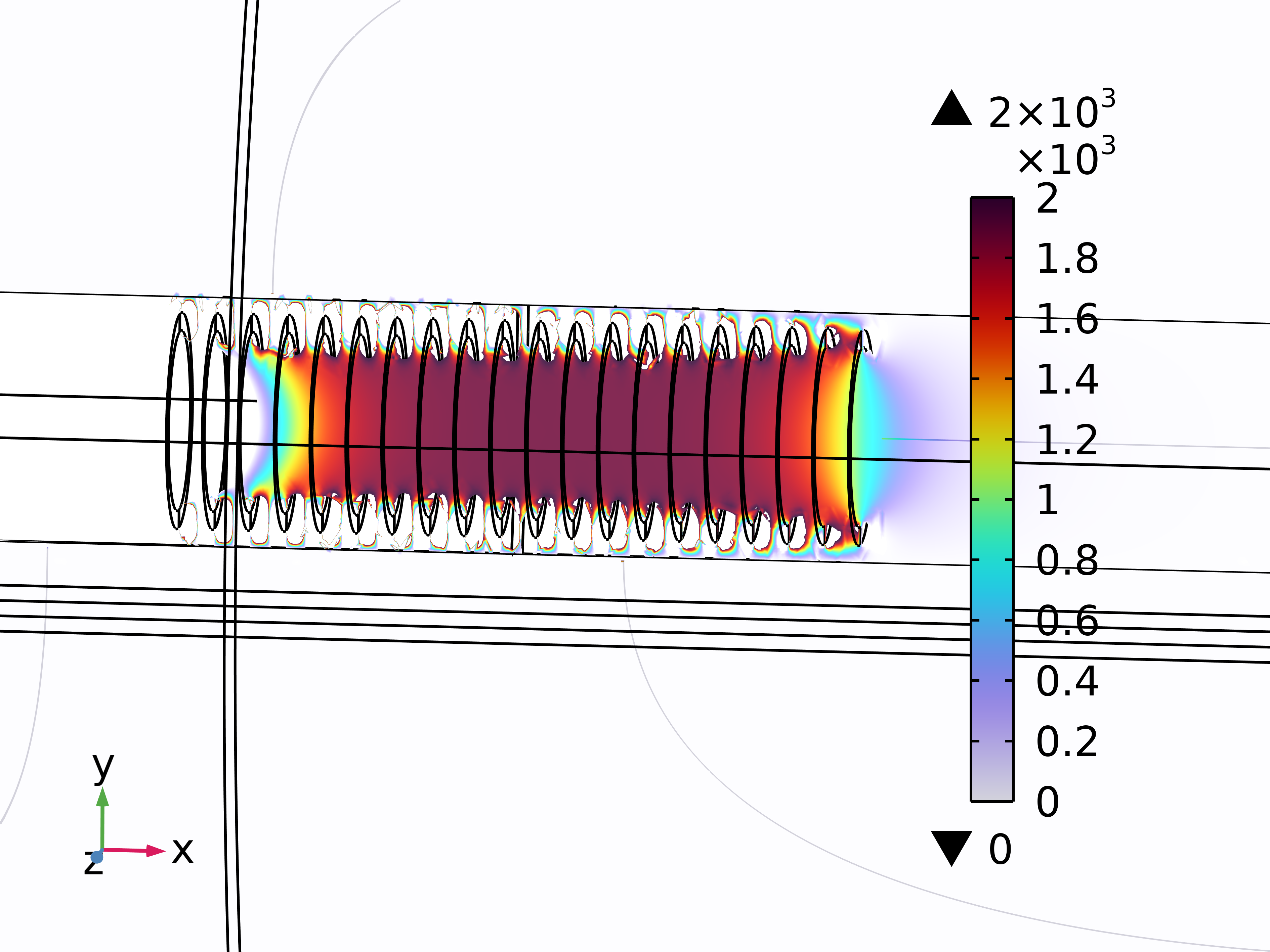}}
\\
\subfloat[The electric potential field distribution in the axial beam direction. The orange-yellow shaded area is the location of the target region, the black line is the electric field in the beam direction, and the red line is the electric potential.]{\includegraphics[width=0.9\linewidth]{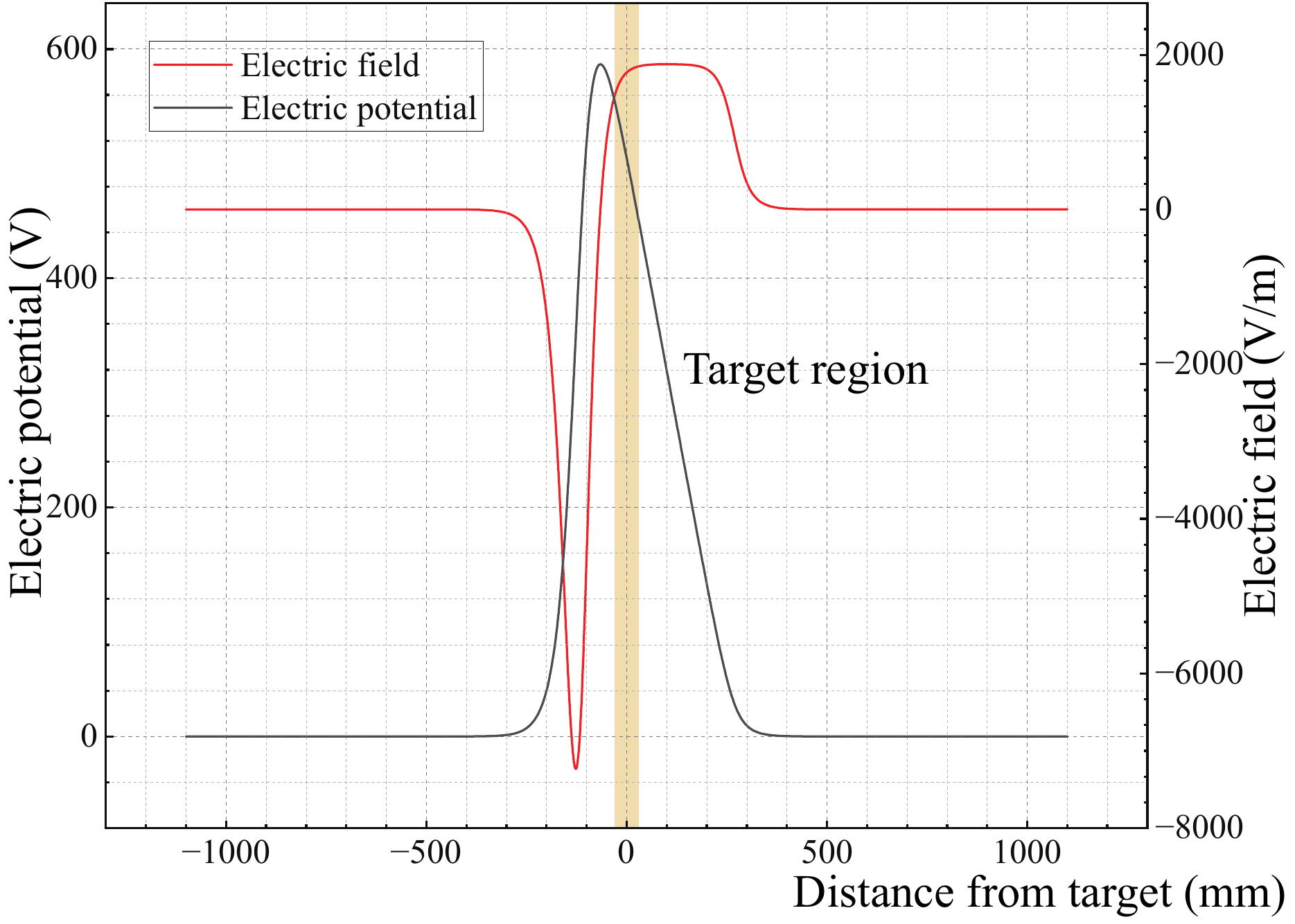}}
\caption{Overview of the electric field effects in the PTS system.}
\label{fig:ele}
\end{figure}

\subsection{Accelerator}

The high muonium yield in vacuum is achieved by injecting surface muons into a silica aerogel target. In this setup, a muon captures an electron in the target to form a thermal muonium. Once a muonium escapes from the target in vacuum, it might convert into an antimuonium, which then decays and produces a positron. The positron carries a low initial momentum at the atomic energy level, a few tens of electronvolts.

To ensure a reasonable signal transmission efficiency, it is necessary to accelerate the positron to at least a few hundred electronvolts. To meet the requirement, we designed an electrostatic accelerator positioned at the center of the detector and integrated within the beam pipe to cover the target region for accelerating signal positrons. To reconstruct the signal with high precision, the system employs multiple circular electrodes. The configuration generates a nearly uniform electric field at the center axis, enabling the accurate reconstruction of positrons from the MCP to the target. A schematic of the accelerator layout and its field distribution is shown in Fig.~\ref{fig:PTS_Overview}.

The accelerator electrodes have an inner diameter of 100~mm and an outer diameter of 120~mm inside of the beam tube of 140~mm. The distance between adjacent plates is 20~mm, resulting in a 41.6~V potential rise between each pair.

\begin{figure*}[!t]
\centering 
\subfloat[At the accelerator exit (VD-a).]{{\includegraphics[width=0.45\linewidth]{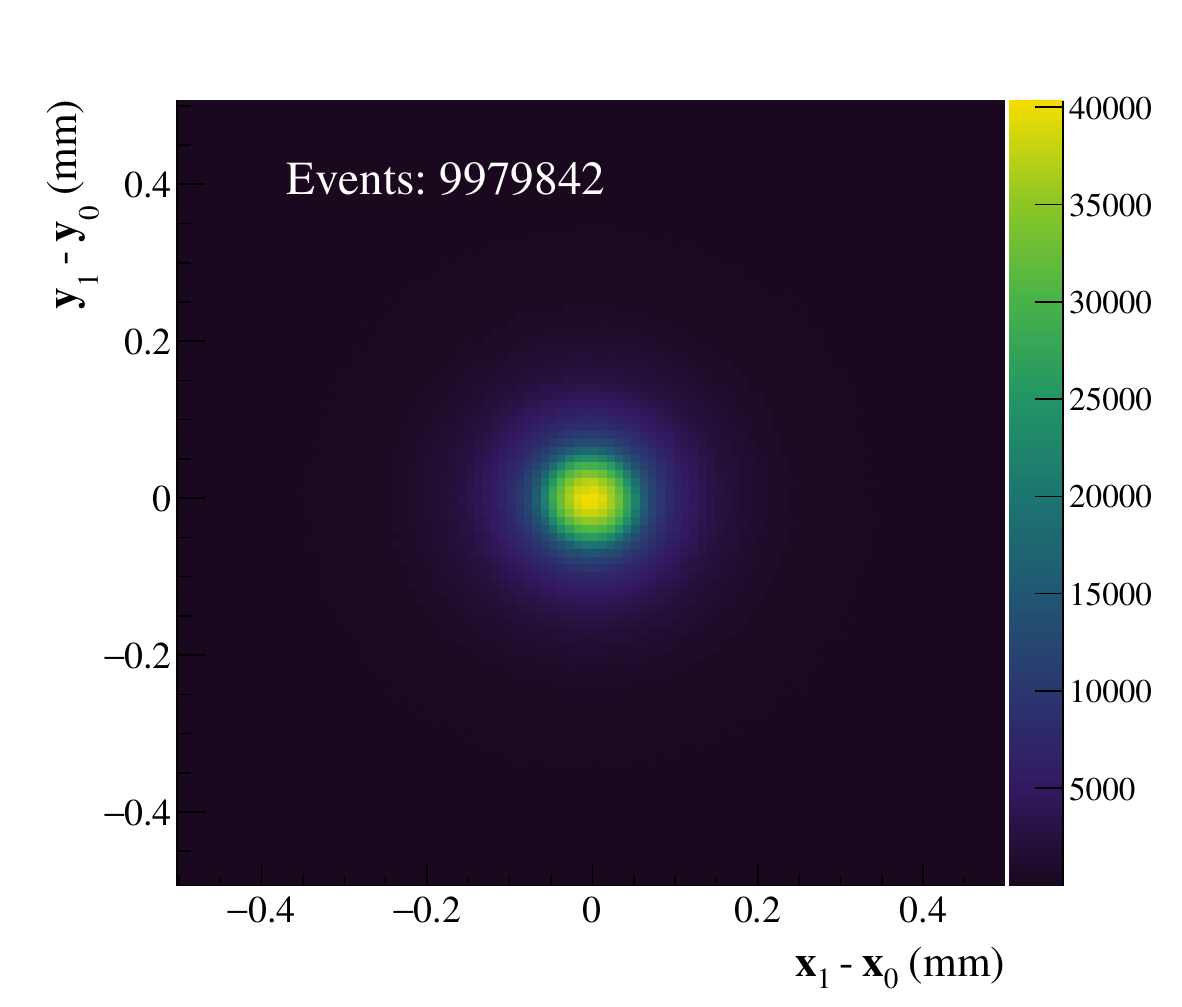}}}
\subfloat[After the first bending solenoid (VD-b).]{{\includegraphics[width=0.45\linewidth]{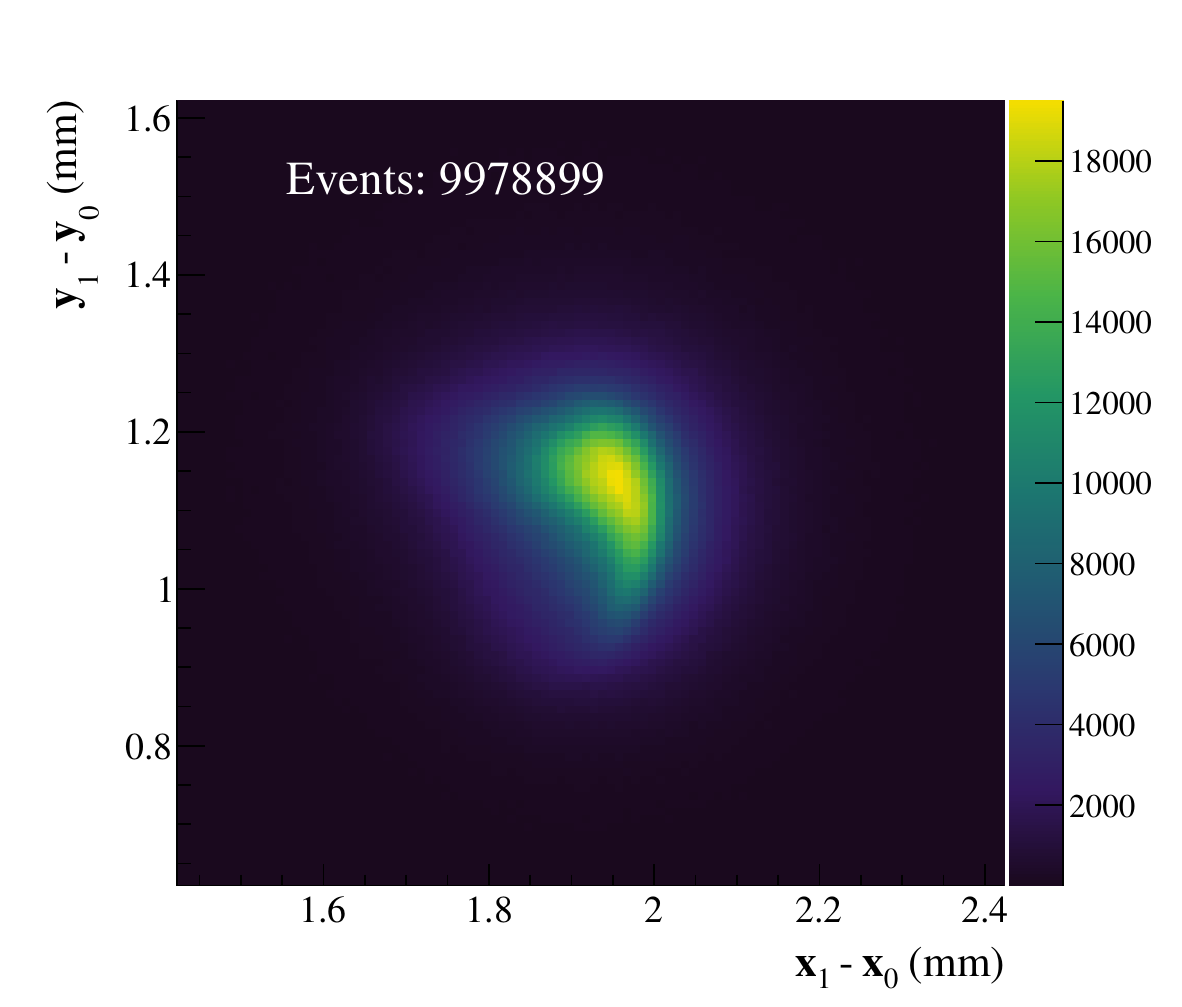}}}
\\
\subfloat[After the collimator (VD-c).]{{\includegraphics[width=0.45\linewidth]{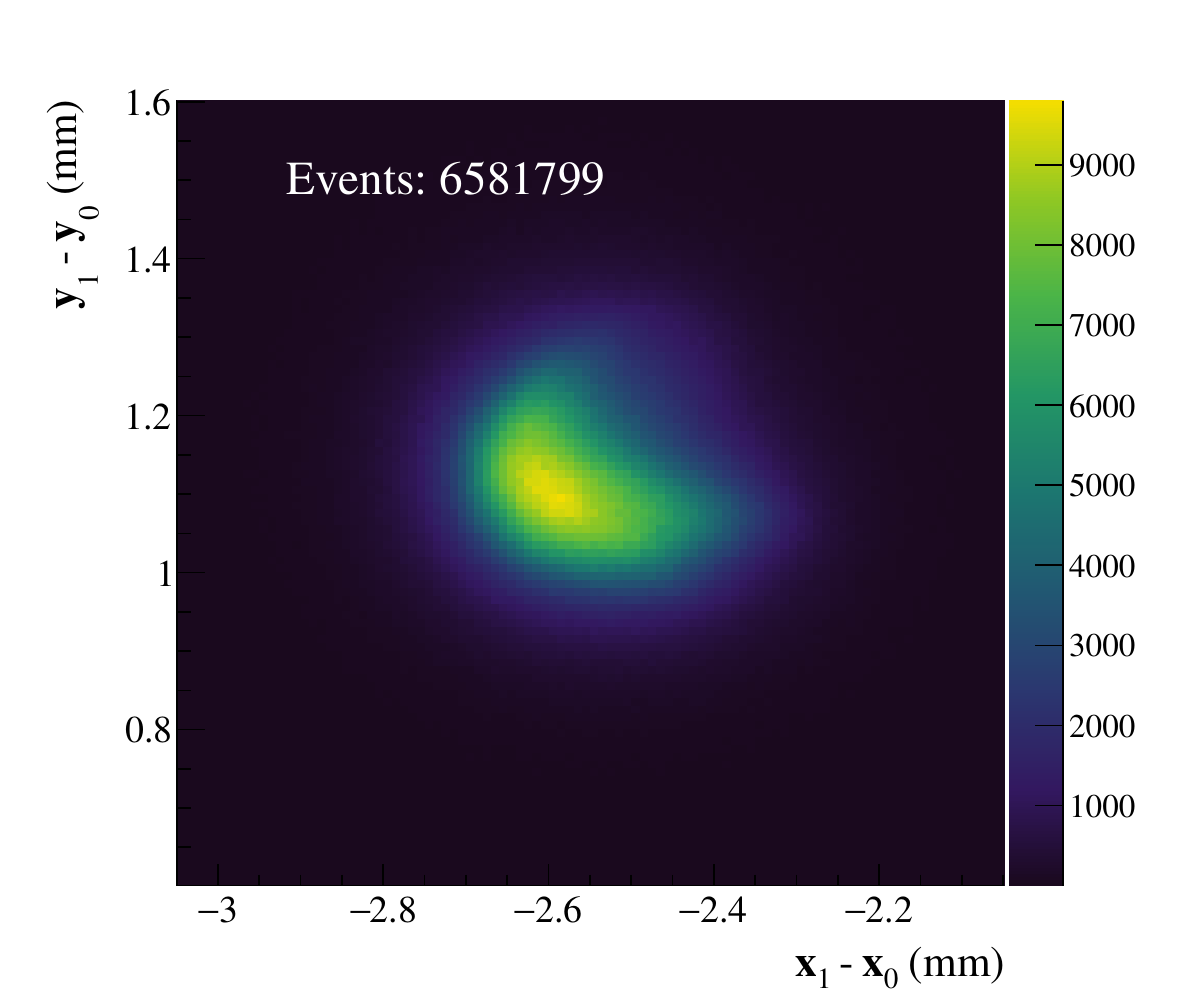}}}
\subfloat[At the PDS center (VD-d).]{{\includegraphics[width=0.45\linewidth]{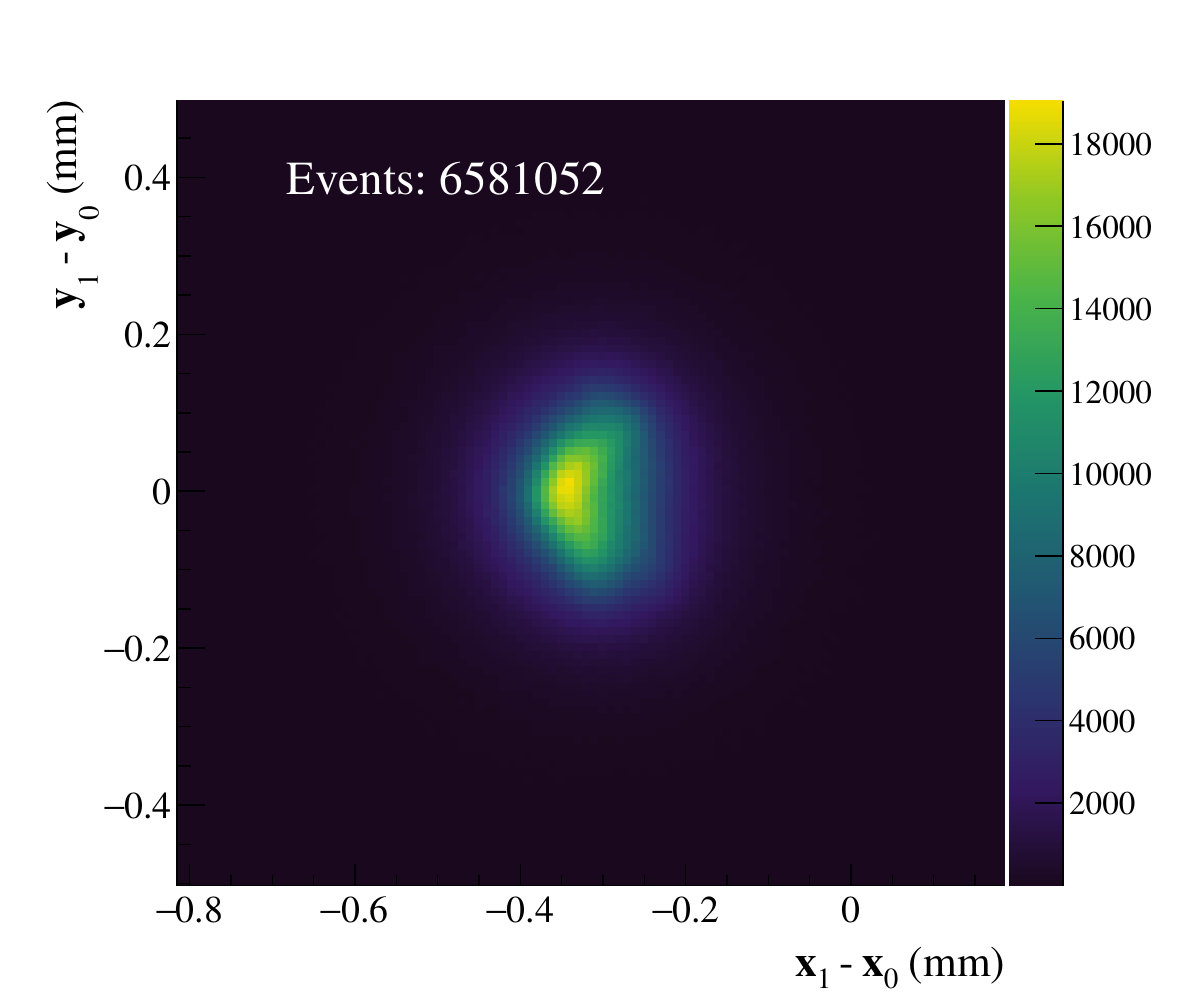}}}
\caption{The relative displacement of the positron. The horizontal and vertical coordinates $x$ and $y$ in the picture are the coordinates recorded on the virtual detector subtracted from the initial coordinates, indicating the offset of the particle during transport, rather than the beam spot. The specific locations of these VDs are indicated in Fig.~\ref{fig:PTS_Overview}.}
\label{fig:PTS_Space_Resolution}
\end{figure*}

For an acceleration module with a total length of 380~mm, the 110~mm distance from the target center to the front end of the accelerator is designed to prevent the target region from falling into the negative electric field region. The downstream length of 270~mm is optimized to minimize the spread of time of flight (TOF), which arises mainly from the initial signal energy of 13.5~eV and the spatial extent of the emission region along the $z$ direction.

The TOF of signal positron can be estimated by the following equation:

\begin{equation}
\begin{aligned}
\text{TOF}=& \frac{-L m v_0 + \sqrt{L m (L m v_0^2 + 2 e U (L - z_0))}}{e U}\\&+ \frac{s - L}{\sqrt{\frac{2 e U (L - z_0)}{L m} + v_0^2}}~,
\end{aligned}
\end{equation}

where TOF represents the time of flight from the antimuonium decay vertex to the hit point on MCP, \( L \) and \( s \) denote the length in the acceleration region and the total flight path, respectively. The initial velocity along the \( z \) direction is \( v_0 \), \( U \) is the acceleration voltage, \( z_0 \) corresponds to the initial downstream offset of the signal source, \( e \) is the positron charge and \( m \) is the positron mass.

For simplicity, the spread of TOF for a signal positron can be characterized by the difference between two typical TOFs, 

\begin{equation}
\Delta \text{TOF} = \text{TOF}_{\text{slow}} -\text{ TOF}_{\text{fast}}~.
\end{equation}
Here, taking a typical kinetic energy of 13.5~eV for positron, $\text{TOF}_{\text{slow}}$ is the longest TOF, corresponding to positrons originating from the downstream side of the acceleration region and emitted opposite to the beam direction, while $\text{TOF}_{\text{fast}}$ is the shortest TOF, corresponding to positrons from the upstream side emitted along the beam direction. By minimizing $\Delta \text{TOF}$, an optimal acceleration length can be found.
After optimization, it is found that with an acceleration potential of 500~V, an acceleration region of 270~mm yields a minimum TOF difference of 24~ns.

The TOF difference can be reduced by increasing the acceleration voltage. For example, at 5000~V, the TOF difference can achieve 4~ns. However, this requires an extended acceleration region of 480~mm and must also consider the optimal detection efficiency range of the MCP, which consequently deteriorates the background rejection capability. Therefore, the current setup adopts such a 500~V acceleration voltage as a compromise.

Fig.~\ref{fig:ele} shows a uniform electric field distribution, except for fringe effects at the front and rear ends. Along the axis of the target region, the electric field strength ranges from 1473 to 1850~V/m, while the potential varies between 450 and 554~V. 

While our simulation results confirm that the uniformity and strength of the electric field meet the physics requirements, significant engineering challenges would arise from installing the accelerator inside the beam pipe and tuning the field. It will be a milestone for MACE to address the practical implementation of PTS in the future.

\section{Simulation results}
\label{sec:results}

\subsection{Signal}
\label{sec:results-a}

The particle transport simulation is performed with the MACE offline software, which is developed based on the Mustard framework and \textsc{Geant4} toolkit~\cite{GEANT4:2002zbu, MustardFramework, MACESW}. By importing the electromagnetic field distributions with \textsc{COMSOL} for the whole system, the framework realizes a full-process simulation covering the entire trajectory.

In the simulation, signal positrons from antimuonium decays traverse the PTS from the target region to the center of the PDS. Fig.~\ref{fig:PTS_Space_Resolution} illustrates the evolution of the transverse displacement distribution of the positron beam at several key locations. We define the geometric acceptance as the probability of the particle passing the PTS. With a total of $10^8$ antimuonium decay events from the initial region in simulation, we can trace the survival event numbers at different locations of the whole detector system, represented by results of virtual detectors in Fig.~\ref{fig:PTS_Space_Resolution}. The panel (d) tells us that we reach a geometric acceptance of 65.81(4)\%. The simulation result is consistent with the estimated collimator signal efficiency given in eq.~(\ref{eq:collimator}), which is the ratio of signal events after the bending magnet and after the collimator in Fig.~\ref{fig:PTS_Space_Resolution} (b) and Fig.~\ref{fig:PTS_Space_Resolution} (c).

Following the initial acceleration, the system achieves  a spatial resolution of $88(1)~\mu\text{m} \times 102(1)~\mu\text{m}$ in the $x$ and $y$ directions, primarily influenced by the initial distribution of the kinetic energy of the positrons and the fringe effects of the electric field. Although traversing the S-shaped solenoid, fringe fields, and path bending temporarily degrade the resolution, the symmetric S-shaped design of the magnetic field effectively compensates for these distortions. In a comparison of the panel (b) and (d) in Fig.~\ref{fig:PTS_Space_Resolution}, the spatial resolution deteriorates after the first bending magnet due to dispersion but is recovered after traversing the second bend, where the spatial resolution improves from $117~\mu\text{m} \times 125~\mu\text{m}$ to $88~\mu\text{m} \times 102~\mu\text{m}$. Furthermore, the symmetric design reduces the drift of the beam center. The mean displacement decreases from $1.9~\text{mm} \times 1.1~\text{mm}$ after the first bend to $0.31~\text{mm} \times 0.0028~\text{mm}$ at the PDS center.

After traversing the entire system, a final geometric acceptance of the signal is achieved at 65.81(4)\%, with positrons arriving at the PDS with a high spatial resolution of 88(1) $\mu$m × 102(1) $\mu$m. In addition to these spatial characteristics, the simulation yields a mean transit time of 322.4(1)~ns with a timing spread (standard deviation) of 6.9(1)~ns. The simulation results for geometric acceptance, precision, and timing characteristics are sufficient to meet the design requirements of MACE.

\subsection{Background}
\label{sec:results-B}

The PTS in MACE plays a crucial role in the discrimination of signals and backgrounds. The PTS provides a direct three-dimensional momentum selection for positrons. The transverse momentum is mainly filtered by the collimator, while longitudinal momentum selection is achieved through a combination of TOF selection and the magnetic rigidity requirements of the S-shaped solenoid.

\begin{figure}[ht!]
    \centering
    \includegraphics[width=1\linewidth]{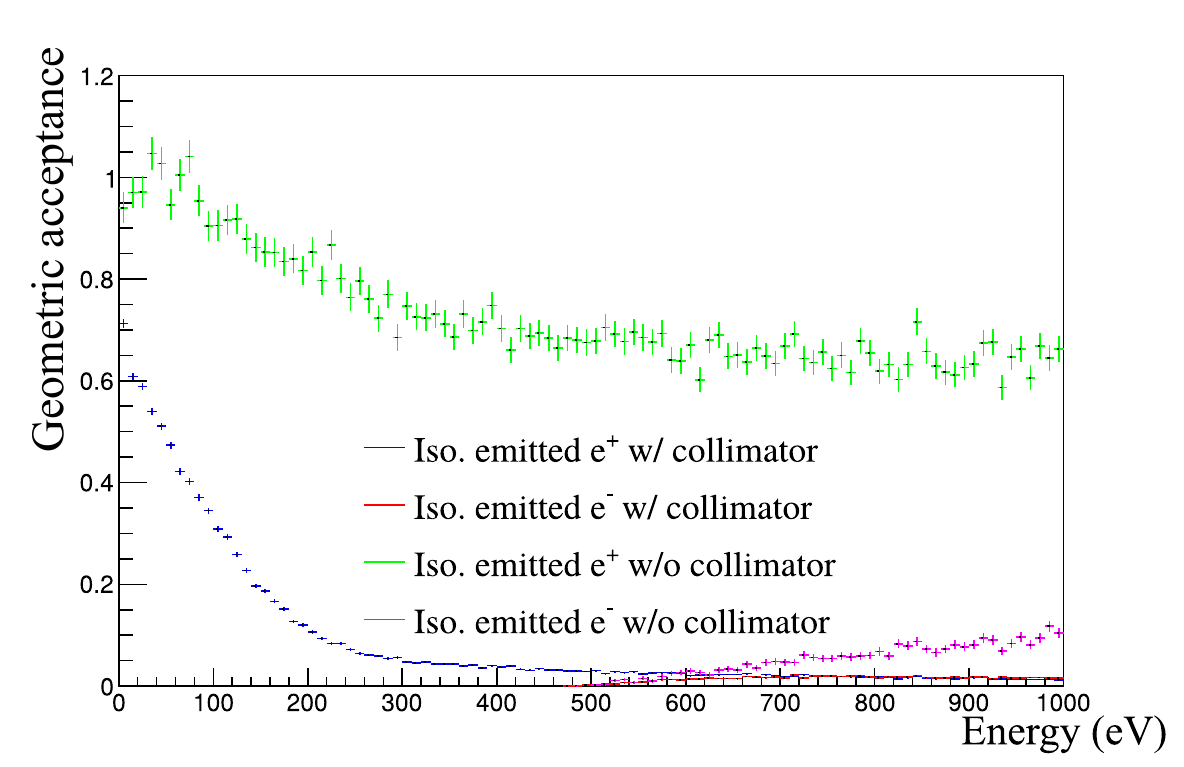}
    \caption{Geometric acceptance of PTS for particles with isotropic initial momentum.}
    \label{fig:PTS_EFF}
\end{figure}

In order to test the capability of PTS to select particles with a certain momentum, we simulate the geometric acceptance curves of $e^{+}$ and $e^{-}$ as a function of kinetic energy. As shown in Fig.~\ref{fig:PTS_EFF}, in the low-energy region (below 1 keV), the geometric acceptance of positrons without the collimator can exceed 60\%, generally reaching approximately 100\% at near-signal energies (below 100~eV). As energy increases beyond 500~eV, the kinetic energy of the electrons can overcome the accelerating electric potential, leading to an increase in geometric acceptance. When a collimator is added, the geometric acceptance in the signal energy region remains around 70\%, consistent with the simulation result in Section \ref{sec:results-a}. With the increase of energy, the loss of geometric acceptance becomes significantly steeper compared to the simulation results without the collimator. At 700~eV, the geometric acceptance drops to the level of \(10^{-2}\).

We also conducted geometric acceptance simulations for various backgrounds, including \( \mu^{+} \), \( \mu^{-} \), \( \pi^{+} \), and \( \pi^{-} \). However, since these particles do not meet the signal identification criteria within the PDS, our primary focus in this study is on backgrounds that can mimic the signal signature, particularly on the IC muon decay process. For a comprehensive analysis of how the PTS can reduce other backgrounds, detailed information can be found in our conceptual design report~\cite{Bai:2024skk}.

The geometric acceptance of positrons from IC muon decay is calculated by multiplying the geometric acceptance by the energy spectrum and integrating over it,
\begin{equation}
\varepsilon_{\text{geom}} = \sum_{\text{bins}} \varepsilon_{\text{geom}}(E_k) \cdot P_{\text{IC}}(E_k)\,,
\label{eq:epsilon_geom}
\end{equation}
where $\varepsilon_{\text{geom}}(E_k)$ is the energy-dependent geometric acceptance derived from simulations (as shown in Fig.~\ref{fig:PTS_EFF}), and $P_{\text{IC}}(E_k)$ represents the normalized probability density of the IC muon decay positrons in each energy bin. As a result, the current system achieves a strong background suppression with a geometric acceptance for IC backgrounds at an extremely low level of $2.1 \times 10^{-6}$.

We further consider the time-of-flight method to enhance background rejection. The total background suppression factor for the IC muon decay is then as follows:
\begin{equation}
\varepsilon_{\text{total}} = \sum_{\text{bins}} \varepsilon_{\text{geom}}(E_k) \cdot \varepsilon_{\text{TOF}}(E_k) \cdot P_{\text{IC}}(E_k)\,.
\label{eq:epsilon_total}
\end{equation}
Here, the term $\varepsilon_{\text{total}}$ is the total background suppression factor, and the $\varepsilon_{\text{TOF}}(E_k)$ denotes the efficiency of TOF cuts for a specific energy bin, corresponding to the selected TOF window.

\begin{figure}[ht!]
    \centering
    \includegraphics[width=1\linewidth]{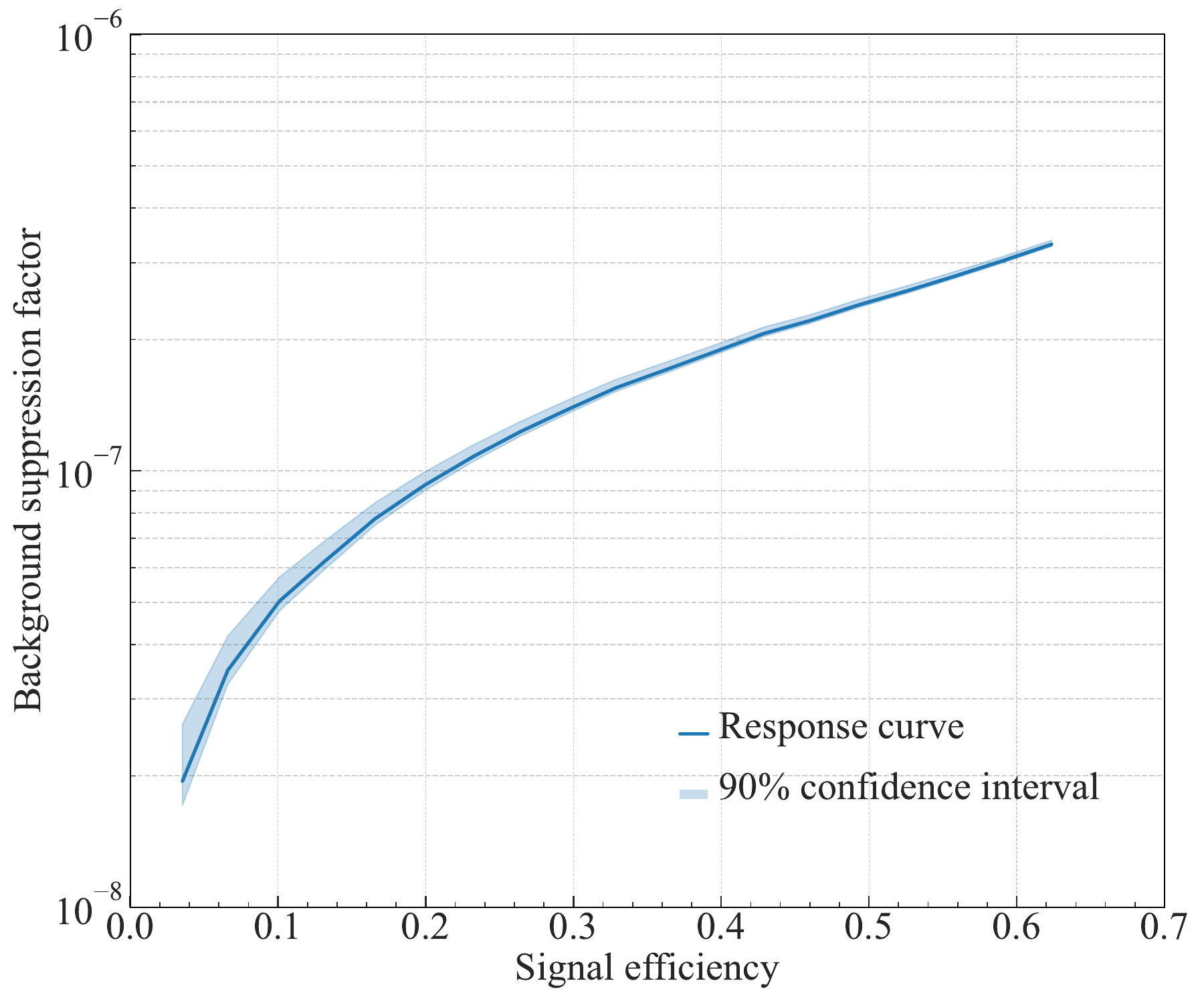}
    \caption{Curve of total efficiencies for signal versus IC background with varying TOF windows. The shaded band indicates the 90\% confidence interval.}
    \label{fig:PTS_ROC}
\end{figure}

Fig.~\ref{fig:PTS_ROC} illustrates the response curves for the total efficiencies of the signal and the background suppression factor as a function of the chosen time window. This curve incorporates statistical errors derived from Monte Carlo simulations of the normalized energy spectra for IC muon decays and background simulations. In the analysis, the TOF window, denoted as $[T_{\text{mean}} - \Delta t, T_{\text{mean}} + \Delta t]$, is given by a symmetric time interval centered on the mean transit time of signal ($T_{\text{mean}} \approx 322.4$~ns). By varying the half-width $\Delta t$ in the analysis, we obtain different signal efficiencies and background rejection factors. By scanning $\Delta t$, we figure out the optimal time window [309.0, 337.9]~ns. Therefore, the signal efficiency is maintained at 62.4\%, while the upper limit for the total rejection factor of a single positron deriving from IC muon decays is determined at $3.0 \times 10^{-7}$. With the current design of PTS, a full simulation of MACE for one year of beam-time, given $10^8~\mu^+$ on target, presents a physical background level of 0.29(2) from IC muon decays~\cite{Bai:2024skk}. The PTS design as a critical component then leads MACE to a background-free experiment.

\section{Conclusion}
\label{sec:conclusion}

In this study, we have presented the design of the PTS for the MACE experiment, achieving the goals of high-precision signal transmission and stringent background suppression. Through comprehensive simulations, the proposed symmetric S-shaped solenoid and the electrostatic accelerator have a spatial resolution of $88(1)\,\mu\text{m} \times 102(1)\,\mu\text{m}$. Besides, the design achieves a total signal efficiency of 62.4\% while suppressing the dominant IC muon decay background by approximately seven orders of magnitude. These results confirm that the PTS design meets the sensitivity requirements of MACE.

Future efforts will be technical design and engineering validation. A primary focus will be the practical implementation of the key components, with particular attention dedicated to installation procedures, mechanical tolerances, and alignment precision of the collimator and solenoid coils. Of critical importance is the integration of the electrostatic accelerator. As an internal part located within the beam pipe, its coupling behavior with both the muonium production target and the surrounding MMS detector requires rigorous investigation to ensure detection performance. Subsequent prototyping and testing will validate the transmission precision of the acceleration and bending segments, laying the solid groundwork for the final experimental success.
\\

\section*{Acknowledgements}
This project is supported in part by
National Natural Science Foundation of China under Grant No. 12075326, Guangdong Basic and Applied Basic Research Foundation under Grant No. 2025A1515010669, Natural Science Foundation of Guangzhou under Grant No. 2024A04J6243, and Fundamental Research Funds for the Central Universities (23xkjc017) in Sun Yat-sen University. J.T. is grateful to Southern Center for Nuclear-Science Theory (SCNT) at Institute of Modern Physics in Chinese Academy of Sciences for hospitality. The simulation benefited greatly from the provision of computing resources by the National Supercomputer Center in Guangzhou. This work was also supported by the National College Students Innovation and Entrepreneurship Training Program, Sun Yat-sen University.

\bibliography{main}

\end{document}